\documentclass[reprint,aps,showpacs,pra,nofootinbib]{revtex4-1}
\usepackage{color}
\usepackage{newlfont}
\usepackage{graphicx}
\usepackage{amssymb}
\usepackage{amsmath}
  
\usepackage{bm}
\usepackage{verbatim}
\usepackage[latin1]{inputenc}
\usepackage{bbm}
\usepackage{multirow}
\usepackage[thinlines]{easytable}
\usepackage{braket}
\usepackage{amsthm}
\usepackage{bbold}
\usepackage{subfig}
\usepackage{mathrsfs}
\usepackage[varg]{txfonts} 
\usepackage{amsfonts}

\newcommand{\be}{\begin{equation}}
\newcommand{\ee}{\end{equation}}
\newcommand{\beq}{\begin{eqnarray}}
\newcommand{\eeq}{\end{eqnarray}}
\newcommand{\tx}{\text}
\newcommand{\mc}{\mathcal}
\newcommand{\mf}{\mathfrak}

\newcommand{\mr}{\mathrm}
\newcommand{\mbb}{\mathbbm}
\newcommand{\mbf}{\mathbf}
\newcommand{\dt}{{\Delta t}}
\newcommand{\sch}{^\mathrm{\mf{s}}}
\newcommand{\cm}{_\mathrm{\text{\tiny cm}}}
\newcommand{\rr}{_\mathrm{\text{\tiny r}}}
\newcommand{\sfrac}[2]{\text{\small $\frac{#1}{#2}$}}
\DeclareMathOperator{\Tr}{Tr}

\begin{document}

\title{Quantum mechanical work}
\author{T. A. B. Pinto Silva}
\author{R. M. Angelo}
\affiliation{Department of Physics, Federal University of Paran\'a, P.O. Box 19044, 81531-980 Curitiba, Paran\'a, Brazil}

\begin{abstract} 
Regarded as one of the most fundamental concepts of classical mechanics and thermodynamics, work has received well-grounded definitions within the quantum framework since the 1970s, having being successfully applied to many contexts. Recent developments on the concept have taken place in the emergent field of quantum thermodynamics, where work is frequently characterized as a stochastic variable. Notwithstanding this remarkable progress, it is still debatable whether some sensible notion of work can be posed for a strictly quantum instance involving a few-particle system prepared in a pure state and abandoned to its closed autonomous dynamics. By treating work as a quantum mechanical observable with a well defined classical limit, here we show that this scenario can be satisfactorily materialized. We prove, by explicit examples, that one can indeed assign eigensystems to work operators. This paves the way for frameworks involving quantum superposition and nonlocal steering of work. We also show that  two-point measurement protocols can be inappropriate to describe work (and other two-time physical quantities), especially in the semiclassical regime. However subtle it may be, our quantum mechanical notion of work is experimentally testable and requires an updating of our intuition regarding the concept of two-time elements of reality. In this context, we derive a work-energy uncertainty relation, and we illustrate how energy conservation emerges as an element of physical reality.  
\end{abstract}

\maketitle

\section{Introduction}
\label{sec:intro}

In classical mechanics, the fundamental law of motion, $m_i\ddot{\mbf{r}}_i=\sum_j\mbf{f}_{j\to i}$, for a point particle of mass $m_i$ at the position $\mbf{r}_i$ submitted to conservative forces $\mbf{f}_{j\to i}=-\nabla_i \mc{V}_{ij}$, where $\mc{V}_{ij}=\mc{V}_{ij}(|\mbf{r}_i-\mbf{r}_j|)=\mc{V}_{ji}$, can be expressed in a form that is, so to speak, ``nonlocal in time'', namely,
\be 
\mc{K}_i(t_2)-\mc{K}_i(t_1)=\sum_j\mc{W}_{j\to i}(t_2,t_1),
\label{DK=W}
\ee 
where $\mc{K}_i(t)=m_i\dot{\mbf{r}}_i^2/2$ and
\be 
\mc{W}_{j\to i}(t_2,t_1)=\int_{t_1}^{t_2}\!\!d t\,\dot{\mbf{r}}_i\cdot\mbf{f}_{j\to i}=-\int_{t_1}^{t_2}\!\!d t\,\dot{\mbf{r}}_i\cdot \nabla_i\mc{V}_{ij}.
\label{Wji}
\ee 
The latter expression defines the mechanical work that particle $j$ does on particle $i$ in the time interval $[t_1,t_2]$ by means of the force $\mbf{f}_{j\to i}$. In addition to providing a useful rephrasing of Newton's second law along a given segment of a configuration space trajectory, the notion of work paves the way for a sensible statement of the law of conservation of energy. In its general form, the principle reads
\be
\Delta \mc{E}_\mr{S}=\mc{W}_\mr{E\to S},
\ee
where $\mr{E}$ stands for ``the environment'' and $\mc{E}_\mr{S}$ for the total energy of ``the system $\mr{S}$'' (the boundary between them being arbitrary and abstract). Clearly, the system energy is conserved only in the absence of external work. In the model discussed above we have $\mr{S} =\{i\}$ and $\mr{E=U-S}$, where $\mr{U}$ denotes the universe set encompassing all particles, so that $\mc{E}_\mr{S}=\mc{K}_i$ and $\mc{W}_\mr{E\to S}=\sum_{j\in{\mr{E}}}\mc{W}_{j\to i}=\int_{t_1}^{t_2}d t\,m_i\,\dot{\mbf{r}}_i\cdot\ddot{\mbf{r}}_i$. When $\mr{S}$ is a many-particle system, the scenario becomes subtler and many definitions of work and related energy changes are admissible even in purely mechanical contexts involving matter systems~\cite{Mallinckro1992,Leff1993}. 

As the focus moves onto huge, messy, dirty, and hot systems, detailed microscopic aspects of the dynamics often become superfluous. In this domain, a proper account for the balance of energy is provided by thermodynamics~\cite{Callen1985,Fermi1936,Zemansky1997}, a phenomenological theory of smooth thermal processes. Supported by the microscopically oriented formalism of statistical physics \cite{Gibbs1902,Tolman1938,Landau1980}, thermodynamics brings into play the notion of heat and defines work in terms of changes of macroscopic degrees of freedom, as for instance the position of a piston. Further adaptations of the notion of work are introduced as one comes to the scope of the so-called stochastic thermodynamics \cite{Sekimoto2010,Deffner2019}, devoted to smaller systems subjected to non-negligible fluctuations. 

Now, how about a genuine quantum mechanical formulation of work? Given the lack of the concept of trajectory in the quantum orthodoxy, which then precludes one from trying a definition such as \eqref{Wji}, one might think at first that any attempt along these lines would be doomed to failure. Despite these difficulties, some interesting developments have been achieved with use of the Heisenberg picture of mechanical systems \cite{Prentis2004}, and, with more intensive efforts, in the emerging field of quantum thermodynamics \cite{Gemmer2005,Millen2016}. In fact, the first attempts to define work in the quantum domain date back to the 1970s (see, e.g., \cite{Alicki1979,Bochkov1977}), and they were devoted to implementing the thermodynamic view according to which work must be related to the transfer of energy through a {\it deterministically controllable} process. These models express the system energy as a Hamiltonian operator $H(\lambda)$ that explicitly depends on a deterministic (dispersion-free) time-dependent parameter $\lambda(t)$ whose dynamics is imposed by an external macroscopic mechanism. Precisely at this point, one recognizes the non-autonomous classical aspect of this model. Adhering to this framework, several approaches on quantum thermodynamics~\cite{Allahverdyan2004,Allahverdyan2005,Talkner2007,Peliti2008a,Sekimoto2010,Roncaglia2014,Ribeiro2016,Francica2017,Hummer2001,Alicki1979,Perarnau2017,Campisi2011,Strasberg2021} compute work as 
\be
\mc{W}(t_2,t_1)=\int_{t_1}^{t_2} \braket{ \partial_t H } dt=\int_{t_1}^{t_2}\braket{\partial_\lambda H}\,\dot{\lambda}\, dt.
\label{W_Alicki}
\ee
In full consonance with the definition of work encountered in statistical physics~\cite{Gibbs1902,Tolman1938}, this formulation is supplemented with a clear recipe for the derivation of a time-dependent Hamiltonian from a closed autonomous dynamics; all one needs to do is to trace out the surroundings and get the operator responsible for the unitary part of the emerging master equation~\cite{Weimer2008,Barra2015,Valente2018,Rodrigues2019}. An important drawback is that this scheme is technically complicated for general strong-coupling regimes. Also, the physical significance of definition~\eqref{W_Alicki} has recently been questioned, since a Hamiltonian like $H(t)+g\mbb{1}$, with an arbitrary real function $g=g(t)$, generates a $g$-independent dynamics but a $g$-dependent amount of work~\cite{Jarzynski2007,Horowitz2007,Vilar2008, Vilar2008a,Vilar2008b,Horowitz2008,Peliti2008,Peliti2008a}. For sure, this cannot be the case in general. Another point under dispute around definition~\eqref{W_Alicki} is whether interaction terms are to be regarded as part of the internal energy of the system. Discussions along these lines have triggered the notions of {\it inclusive} and {\it exclusive} works \cite{Jarzynski2007}. 

There are many other routes to the definition of thermodynamic work in quantum regime, among which we refer the reader to  \cite{Tonner2005,Uzdin2015,Lorch2018,Skrzypczyk2014,Binder2015,Alipour2016,Miller2017,Valente2021a,Valente2021b,Deffner2016}. In contrast, the literature still lacks a proper account for the definition of work in genuine quantum {\it mechanical} scenarios. Incidentally, it is clear that the knowledge accumulated so far, however useful it may be in the thermodynamics context, does not apply to these cases. Consider, for instance, a particle of mass $m$ (the system, $\mr{S}$) prepared in a pure state and submitted to some external influence (the environment, $\mr{E}$) described by a harmonic potential $V(X)=kX^2/2$. Since the quantum dynamics is determined by the parameter-free Hamiltonian $H=P^2/2m+kX^2/2$, a direct application of equation \eqref{W_Alicki} would give $\partial_\lambda(P^2/2m)=0$ and hence no work at all. This is, however, unexpected because the average kinetic energy of the particle surely varies over time. As a second example, let us take an isolated quantum system composed of two spinless particles. Let $H=P_1^2/2m_1+P_2^2/2m_2+V_{12}(|X_1-X_2|)$ be the Hamiltonian governing the autonomous dynamics, and let $\rho_0$ be a correlated joint state without any connection {\it a priori} with thermal states. Even if one is able to derive a master equation for an arbitrary-strength potential $V_{12}$ and eventually find an effective driving parameter $\lambda(t)$, the resulting effective Hamiltonian for the reduced dynamics will not be free from the issue of encompassing part of the coupling and, as consequence, the work \eqref{W_Alicki} will be submitted to the inclusive-versus-exclusive polemic. In addition, this approach does not shed light on the fundamental question of whether work can be treated as an observable and how one can measure it.

This article aims to introduce a genuine quantum mechanical notion of work, by which we mean a concept applicable even to the problem of a single quantum particle prepared in a pure state and submitted to a scalar time-independent potential. The text is structured as follows. Drawing some inspiration from the classical formalism, which is briefly reviewed in the beginning of Sec.~\ref{sec:MechWork}, we define a Hermitian quantum work operator that avoids, by construction, the inclusive-versus-exclusive dilemma. In Sec.~\ref{sec:observable}, we explicitly obtain the quantum work eigensystem for the problems of a particle submitted to a uniform gravitational field (Sec.~\ref{sec:gravity}) and two particles under a quadratic coupling (Sec.~\ref{sec:quadratic}). The work statistics corresponding to the gravitational problem is derived in Sec. \ref{sec:statistics}, where a comparison is made with the results offered by the stochastic-variable approach. We show that the mean work correctly reproduces the semiclassical results expected in light of the Ehrenfest theorem. In addition, we discuss how work and other two-time quantities can be interpreted as elements of the physical reality (Sec.~\ref{sec:EoR}), we show that a Schr\"odinger-like picture is admissible for work (Sec.~\ref{sec:SchPict}), and we derive a work-energy uncertainty relation (Sec.~\ref{sec:uncertainty}). In particular, we illustrate how energy conservation emerges as an element of reality. Section~\ref{sec:conclusions} then closes the paper with our concluding remarks.

\section{Mechanical work}
\label{sec:MechWork}

Before presenting our definition of quantum mechanical work, it is instructive to briefly review an important property of the mechanical work \eqref{Wji} in systems of structureless particles. From now on, we confine our analysis to one-dimensional motion, for the sake of simplicity. Consider the Hamiltonian function
\be
\mc{H}=\sum_{i=1}^N\frac{p_{i}^{2}}{2m_{i}}+\sum_{j> i=1}^N\mc{V}_{ij},
\label{HN}
\ee 
with $\mc{V}_{ij}\equiv \mc{V}_{ij}(|x_i-x_j|)$. The notation is such that the second parcel above encompasses summations over $i$ and $j>i$. Assume, for a while, that $N=3$ and select particle 2 as the system of interest ${\mr{S}}$, so that the environment ${\mr{E}}$ is composed of particles 1 and 3. In this case, from $\mc{W}_{j\to i}=-\int_{t_1}^{t_2}dt\,\dot{x}_i\,\partial_{x_i}\mc{V}_{ij}$ and the chain rule we find
\beq 
\mc{W}_\mr{E\to S}&\equiv& \mc{W}_{1\to 2}+\mc{W}_{3\to 2}=\int_{t_1}^{t_2}dt\,\dot{x}_{2}\, m\ddot{x}_{2} \nonumber \\  &=& \mc{K}_{2}(t_2)-\mc{K}_2(t_1) \equiv \Delta \mc{E}_{\mr{S}}.
\label{DK2}
\eeq 
Thus, departing from the (unambiguous) notion of external work $\mc{W}_\mr{E\to S}$, we arrive at the notion of internal energy $\mc{E}_\mr{S}$, which here is the kinetic energy of $\mr{S}$. Notice that the interaction terms $\mc{V}_{12}$ and $\mc{V}_{23}$ do not contribute to the internal energy. This is reasonable because they do not even exist when the particle 2 is left alone, so they cannot be ``internal''. Let us now redefine the abstract boundary between system and environment: $\mr{S}=\{1,2\}$ and $\mr{E}=\{3\}$. With use of the identity $\frac{d }{dt}(\mc{K}_1+\mc{K}_2+\mc{V}_{12})=-\dot{x}_1\partial_{x_1}\mc{V}_{13}-\dot{x}_2\partial_{x_2}\mc{V}_{23}$ and the previous relation for $\mc{W}_{j\to i}$, one shows that
\be
\mc{W}_\mr{E\to S}\equiv \mc{W}_{3\to 1}+\mc{W}_{3\to 2}=\Delta\Big(\mc{K}_{1}+\mc{K}_{2}+\mc{V}_{12}\Big)\equiv \Delta \mc{E}_\mr{S}.
\label{W3-12}
\ee
Again, the result is intuitive: the internal energy $\mc{E}_\mr{S}=\mc{K}_1+\mc{K}_2+\mc{V}_{12}$ now includes the interaction $\mc{V}_{12}$ taking place inside the boundary that defines the system. Naturally, the interaction terms $\mc{V}_{13}$ and $\mc{V}_{23}$ are not part of the internal energy, since they cannot exist in the system $\mr{S}=\{1,2\}$ alone. Here is the point we want to emphasize: once the mechanical work is properly defined, the correct notion of internal energy emerges automatically and the inclusive-versus-exclusive debate dissipates. The extension to arbitrary $N$ is straightforward. Indeed, for $\mr{S}=\{1,2,\cdots,M\}$ and $\mr{E}=\{M+1, M+2,\cdots,N\}$, we find
\be
\mc{W}_\mr{E\to S}=\sum_{j\in \mr{E}}\sum_{i\in \mr{S}}\mathcal{W}_{j\rightarrow i}=\Delta\left(\sum_{i=1}^{M}\mc{K}_i+\sum_{k>i=1}^{M}\mc{V}_{ik}\right)\equiv \Delta \mc{E}_\mr{S}.
\label{WE-N}
\ee
Therefore, in this fundamental, microscopic, mechanical, conservative framework, the principle of conservation of energy is trivially proved, no ambiguity arises concerning the notion of internal energy, and there is no need for one to conceive either uncontrollable or inaccessible forms of energy transfer, such as heat. 

\subsection{Quantum mechanical work}
\label{sec:QMWork}

We are now ready to introduce our definition of quantum mechanical work. The proposal consists of closely following the classical structure \eqref{Wji}. An important obstacle in this regard might be the absence of direct notions of velocity and force in the quantum domain, but as we show next, this can be remedied with use of the Heisenberg picture, wherein the operators are written as $O\equiv O(t)\equiv \phi_t(O\sch)$, where $O\sch$ is the corresponding Schr\"odinger operator\footnote{The superscript $\mf{s}$ will be used hereafter to denote operators in the Schr\"odinger picture. This notation was not employed in Sec.~\ref{sec:intro}, where all operators were implicitly assumed to be written in this picture.} 
and $\phi_t$ is the time-evolution map satisfying $\phi_t^*=\phi_{-t}$ and $\phi_{t_1}\phi_{t_2}=\phi_{t_1+t_2}$. In particular, for unitary evolutions respecting $i\hbar\,\dot{U}_t=HU_t$, one has $\phi_t^*(\rho_0)=U_t\,\rho_0\,U_t^\dag$. Let us consider a universe ${\mr{U}}=\{1,2,\cdots,N\}$  composed of $N$ spinless interacting particles whose dynamics is described by the Hamiltonian operator
\be
H=\sum_{i=1}^N\sfrac{P_i^{2}}{2m_i}+\sum_{j>i=1}^NV_{ij},
\label{HN}
\ee 
where $V_{ij}=V_{ij}(|X_i-X_j|)$. In Heisenberg's picture, the velocity operator and the resultant force operator for the $i$-th particle are respectively written as $\dot{X}_i=[X_i,H]/i\hbar$ and $m_i\ddot{X}_i=m_i[\dot{X}_i,H]/i\hbar$. The quantum state $\rho_0$ and the Hamiltonian $H$ act on a joint Hilbert space $\mbb{H}_{\mr{U}}=\bigotimes_{k=1}^N\mbb{H}_k$, whereas $X_i$ and $P_i$ act on $\mbb{H}_i$. We then introduce our candidate for the {\it resultant quantum mechanical work} done on particle $i$:
\be
W_\mr{E\to S}(t_2,t_1)=\tfrac{1}{2}\int_{t_1}^{t_2}\!\!dt\,\,m_i\{\dot{X}_i,\ddot{X}_i\}
\label{RQMW}, 
\ee
where ${\mr{S}}=\{i\}$ and ${\mr{E}}={\mr{U}-\mr{S}}$. The symmetrization $\frac{1}{2}\{A,B\}=\frac{1}{2}(AB+BA)$ aims at ensuring Hermiticity. It should be recognized from the very beginning that $W_\mr{E\to S}$ is an operator acting on the joint Hilbert space $\mbb{H}_{\mr{U}}=\mbb{H}_{\mr{S}}\otimes\mbb{H}_{\mr{E}}$, the consequences of which  will be discussed posteriorly. Now, one can show by direct integration of the Heisenberg equation $d\dot{X}_i^2/dt=[\dot{X}_i^2,H]/i\hbar=\{\dot{X}_i,\ddot{X}_i\}$ that
\be 
\frac{m_i\dot{X}_i^2(t_2)}{2}-\frac{m_i\dot{X}_i^2(t_1)}{2}=\frac{1}{2}\int_{t_1}^{t_2}\!\!dt\,\,m_i\{\dot{X}_i,\ddot{X}_i\},
\label{QM-DK=W}
\ee 
which is the algebraic statement of the energy-work theorem $\Delta E_{\mr{S}}=W_\mr{E\to S}$ with respect to the internal energy operator $E_{\mr{S}}\equiv K_i(t)=m_i\dot{X}_i^2(t)/2=P_i^2/2m_i$. Also, just as in the classical context, this approach allows us to speak of the notion of ``work done by a force''. Plugging the Heisenberg equation $m_i\ddot{X}_i=\dot{P}_i=-\sum_{j\in {\mr{E}}}\partial_{X_i}V_{ij}$ into \eqref{RQMW} induces us to introduce
\be 
W_{j\to i}(t_2,t_1)=-\text{\small $\frac{1}{2}$}\int_{t_1}^{t_2}\!\!dt\,\{\dot{X}_i,\partial_{X_i}V_{ij}\},
\label{QM-Wji}
\ee 
which makes direct reference to its classical counterpart  \eqref{Wji}. This means that whenever one is able to identify a specific physical potential in the system under consideration, then the recipe \eqref{QM-Wji} can be used to describe the work mediated by the interaction $V_{ij}$ in the time interval $[t_1,t_2]$. The sum of all individual contributions yields the resultant quantum mechanical work, $W_\mr{E\to S}=\sum_{j\in{\mr{E}}}W_{j\to i}$, which causes the change in the internal energy, as prescribed by \eqref{QM-DK=W}. The quantum counterpart of \eqref{WE-N} can be obtained via calculations similar to those employed around \eqref{W3-12}. Setting ${\mr{S}}=\{1,2,\cdots,M\}$ and ${\mr{E}}=\{M+1,M+2,\cdots,N\}$ one shows via some algebra that
\be
W_\mr{E\to S}=\sum_{j\in {\mr{E}}}\sum_{i\in {\mr{S}}}W_{j\rightarrow i}=\Delta\left(\sum_{i=1}^{M}K_i+\sum_{k>i=1}^M V_{ik}\right)\equiv \Delta E_{\mr{S}}.
\label{QM-W=DE}
\ee
Hence, in perfect analogy with the classical framework, the definition of the total work imparted on the system ${\mr{S}}$ reveals the internal-energy operator $E_{\mr{S}}=H_{\mr{S}}$ without any ambiguity. In particular, it is notorious that the Schr\"odinger version $E_{\mr{S}}\sch=H_{\mr{S}}\sch\otimes\mbb{1}_{\mr{E}}$ of this operator effectively acts on $\mbb{H}_{\mr{S}}$ only, and so does $E_{\mr{S}}$ when one ``turns off'' the interaction with the environment. That is, the coupling between ${\mr{S}}$ and ${\mr{E}}$ does not count as internal energy. The theoretical strategy of ``turning off'' the environment allows us to identify what the internal energy should be in our formalism, thus avoiding the inclusive-versus-exclusive ambiguity.

It is opportune to remark that our approach can be straightforwardly extended to more than one dimension, thus being able to implement the principle of conservation of energy, $\Delta E_{\mr{S}}=W_\mr{E\to S}$, in general mechanical contexts. In scenarios involving electromagnetic phenomena, if one can unambiguously identify the internal energy operator $E_{\mr{S}}$ (for instance by ``turning off'' all interactions with the environment), then one can take \eqref{QM-W=DE} as a fundamental postulate, from which the work done on ${\mr{S}}$ by ${\mr{E}}$ can be computed as $\Delta E_{\mr{S}}$. For instance, if the internal energy is represented by some Hamiltonian operator $H_{\mr{S}}\sch\in\mbb{H}_{\mr{S}}$, then the relation
\be 
W_\mr{E\to S}(t_2,t_1)=\phi_{t_2}\left(H_{\mr{S}}\sch\otimes\mbb{1}_{\mr{E}}\right)-\phi_{t_1}\left(H_{\mr{S}}\sch\otimes\mbb{1}_{\mr{E}}\right)
\label{phi(dE)} 
\ee 
gives the work operator done on ${\mr{S}}$ by ${\mr{E}}$ in the time interval $[t_1,t_2]$. In our approach, generic expectation values are computed as $\braket{f(W_\mr{E\to S})}(t_2,t_1)=\Tr[f(W_\mr{E\to S})\rho_0]$, for well behaved functions $f$. In particular, the mean work and its uncertainty follow from
\begin{subequations}
\beq 
&&\braket{W_\mr{E\to S}}=\Tr\left(W_\mr{E\to S}\,\rho_0 \right),\label{<W>}\\
&&\sigma_{W_\mr{E\to S}}=\left(\braket{W_{\mr{E} \to S}^2}-\braket{W_{\mr{E} \to S}}^2\right)^{1/2}, \label{sigma_W}
\eeq 
\end{subequations}
with analogous relations for the work imparted by a specific force [Eq. \eqref{QM-Wji}]. Perhaps the first sanity check one may require from any candidate of quantum mean work is the capability of correctly retrieving its classical counterpart for sufficiently narrow wave packets or less-than-cubic potentials. As we shall explicitly show later, this is indeed the case.

\section{Work as an observable}
\label{sec:observable}

Technically, an {\it observable} is a Hermitian operator whose set of eigenvectors spans the vector space. It then follows from the quantum axioms that as soon as a measurement is concluded, the system state collapses to one of the eigenstates of the measured observable and an associated eigenvalue is obtained as  outcome. Keeping this concept in mind, we come to an important point underlying our approach. Although there are proposals defending the idea that work can indeed be viewed as an observable \cite{Allahverdyan2005,Beyer2020,Gelin2008,Lindblad1983}, such an idea has been an object of intense discussion and it remains unsettled \cite{Miller2017,Talkner2007,Allahverdyan2014,Campisi2011,Talkner2016}.  Some authors argue that because work is related to a process rather than a system state, a single measurement at a given instant of time cannot completely characterize it \cite{Campisi2011}. In another vein \cite{Hanggi2017,Talkner2007,Perarnau2017}, it has been shown that taking {\it fluctuation theorems}\footnote{Fluctuation theorems have shown to be ubiquitous tools leading to several developments in the scope of quantum thermodynamics \cite{Hanggi2017,Campisi2011}.} as fundamental premises favors the view that work is a random variable accessible via two-point measurements (TPM)~\cite{Talkner2007}, to the detriment of  operator-based formulations, like $W_\mr{E\to S}(t_2,t_1)=H_{\mr{S}}(t_2)-H_{\mr{S}}(t_1)$. However, despite the demonstrated relevance of fluctuation theorems in classical stochastic thermodynamics, they are not expected to be valid in general quantum contexts, so that a work operator should not be automatically discarded for not leading to the usual fluctuations theorems, as pointed out in Ref. \cite{Allahverdyan2014}. On the other hand, in the same reference a scenario is presented in which a system is prepared in an eigenstate of work while the energies at different instants of time are uncertain. This led the author to the odd conclusion that energy exchanges and work cannot be related in such a case. It is noteworthy that, to the best of our knowledge, all this debate has been conducted exclusively within the framework of the (quantum and classical) stochastic thermodynamics. We remind the reader that our goal here is to assess the notion of quantum work within a mechanical perspective which, of course, may not necessarily be immune to the aforementioned objections. Moreover, we aim at fundamentally preserving the axiom according to which {\it every} physical quantity is to be described within the quantum formalism as an observable. In what follows, we discuss two emblematic physical problems through which we demonstrate the adequacy of this perspective.

\subsection{Work due to a uniform gravitational field}
\label{sec:gravity}

We start with the simple instance in which a particle ${\mr{S}}$ of mass $m$ is immersed in a homogeneous gravitational field $g$ created by a massive body ${\mr{E}}$ (the Earth), which remains fixed at the origin of the inertial coordinate system. This turns out to effectively be a one-body model described by
\be
H=\frac{P^{2}}{2m}+mgX,
\label{Hgrav}
\ee
where $X$ ($P$) stands for the position (momentum) operator.  We believe that nobody would object to the idea that ${\mr{E}}$ does some work on ${\mr{S}}$, and yet, since there is no external driving $\lambda(t)$, prescription \eqref{W_Alicki} yields $\partial_\lambda(P^2/2m)=0$ and $\mc{W}(t_2,t_1)=0$. To apply our formalism, we first compute the Heisenberg equations $\dot{X}=P/m$ and $\dot{P}=-mg=m\ddot{X}$, and then we find the solutions $P(t)=P\sch-mg t=m\dot{X}$ and $X(t)=X\sch+P\sch t/m-gt^2/2$. Using \eqref{RQMW} and \eqref{QM-DK=W}, we obtain the work operator
\be 
W_\mr{E\to S}(t_2,t_1)=\Delta E_{\mr{S}}=-g\left(t_2-t_1\right)\,P\sch+\tfrac{mg^2}{2}\left(t_2^2-t_1^2\right),
\label{Wg-DK}
\ee 
where $E_{\mr{S}}=K=m\dot{X}^{2}/2$. Clearly, the potential $V=mgX$ is not part of the internal energy of ${\mr{S}}$. Since the operators of the (classical) heavy particle ${\mr{E}}$ do not enter the model, then $\frac{1}{2}dt \{\dot{X},\partial_XV\}=\dot{V}dt$. It follows from \eqref{QM-Wji} that
\be 
W_\mr{E\to S}(t_2,t_1)=-\Delta V=mg\left[X(t_1)-X(t_2)\right].
\label{Wg-DV}
\ee 
Not surprisingly, Eqs. \eqref{Wg-DK} and \eqref{Wg-DV} give $\Delta (K+V)=\Delta H=0$. It is clear from the above results that the work operator has the same right as $P\sch$ to be understood as an observable. In particular, it is Hermitian and satisfies the relation
\be
W_\mr{E\to S}(t_2,t_1)\ket{p}=w_p(t_2,t_1)\ket{p},
\label{Eig-grav}
\ee
with eigenvalues $w_p(t_2,t_1)=-g(t_2-t_1)p+mg^{2}\left(t_2^2-t_1^2\right)/2$ and eigenvectors $\ket{p}$ such that $P\sch\ket{p}=p\ket{p}$. 

As an interesting consequence of Eq. \eqref{Eig-grav}, by measuring the momentum of the particle and thus preparing, say, $\rho_0=\ket{p_0}\bra{p_0}$ as an initial state, one automatically prepares the amount $w_{p_0}(t_2,t_1)$ of work that ${\mr{E}}$ will impart on ${\mr{S}}$ in the time interval $[t_1,t_2]$. Since the initial state is a work eigenstate, one has a scenario in which there is no uncertainty whatsoever for the observable work. Also, it follows that $\braket{W_\mr{E\to S}}=w_p(t_2,t_1)$, which coincides with the value expected from classical mechanics. For instance, we have $w_{p_0}(t,0)={\mr{f}}_g\, \Delta x(t)$, where ${\mr{f}}_g=-mg$ and $\Delta x(t)=p_0t/m-gt^2/2$. This is in full agreement with the Ehrenfest theorem. Remarkably, through a single measurement, one is able to set the amount of work that will be executed in an arbitrary time interval. This is not to say, though, that work is a (time local) state variable. In fact, by construction, the work observable is a two-time operator satisfying $W_\mr{E\to S}(t,t)=0$. In addition, it is clear that the resulting eigenvalue makes reference to a classical trajectory whose segment $x(t_2)-x(t_1)$ characterizes the process through which ${\mr{E}}$ changes ${\mr{S}}$'s internal energy. Notice that the aforementioned preparation scheme applies to other two-time Heisenberg operators as well. Consider, for instance, the displacement operator $\delta(t_2,t_1):= X(t_2)-X(t_1)$ for a free particle. Solving the Heisenberg equations, one straightforwardly finds $\delta(t_2,t_1)=P\sch(t_2-t_1)/m$. We see that by preparing a momentum eigenstate, we can set the displacement of the particle for any future time interval (see Appendix \ref{sec:displacement} for further discussions regarding this observable).

To avoid issues concerning the normalization of the quantum state, one can always associate $\rho_0=\ket{p_0}\bra{p_0}$ with a very sharp Gaussian state or think of it as the projection $\rho_0=\int_{p_0-d_p/2}^{p_0+d_p/2}dp\,\ket{p}\bra{p}\cong \ket{p_0}\bra{p_0}d_p$, with $d_p$ being the resolution for momentum measurements. Alternatively, we can use methods that treat momentum as a discrete variable \cite{Freire2019}. In any of theses cases, however, from a physical point of view, we do expect some superposition of momentum eigenvectors, which naturally implies a superposition of work, with an associated mean value and quantum uncertainty.

\subsection{Work due to an elastic potential}
\label{sec:quadratic}

Now we conceive a universe ${\mr{U}}=\{1,2\}$ with two particles of masses $m_{1,2}$ interacting via an elastic potential of characteristic constant $k$. The unitary autonomous dynamics is governed by the Hamiltonian
\be
H=\frac{P_{1}^{2}}{2m_{1}}+\frac{P_{2}^{2}}{2m_{2}}+\frac{k}{2}(X_{2}-X_{1})^{2},
\label{H-elastic}
\ee
where $X_i$ ($P_i$) is the position (momentum) operator of the $i$-th particle. Here we set ${\mr{S}}=1$ and ${\mr{E}}=2$. By means of the usual transformation
\be
\begin{array}{lll}
	\displaystyle X\cm=\frac{m_{1}X_{1}+m_{2}X_{2}}{M},&\quad &\displaystyle X\rr=X_{2}-X_{1},\\ \displaystyle	P\cm=P_{1}+P_{2},& &\displaystyle P\rr=\mu\left(\frac{P_2}{m_2}-\frac{P_1}{m_1}\right),
\end{array}\label{eq:relacm}  
\ee
with $\mu=m_{1}m_{2}/M$ and $M=m_{1}+m_{2}$, we can rewrite the Hamiltonian in the form $H=P\cm^2/2M+P\rr^2/2\mu+kX\rr^2/2$, 
which shows that the center-of-mass part decouples from the relative one. As a consequence, one can write the time-evolution operator as $U_t=U\cm\otimes U\rr$, which then yields the solutions 
\be
\begin{array}{lll}
	X\cm=X\cm\sch+\frac{P\cm\sch t}{M}, & &  X\rr=X\rr\sch\cos\left(\omega t\right)+\frac{P\rr\sch}{\mu\omega}\sin\left(\omega t\right), \\  \\
	P\cm=P\cm\sch,	& &
	P\rr=P\rr\sch\cos\left(\omega t\right)-\mu\omega  X\rr\sch\sin\left(\omega t\right),
\end{array}
\ee
with $\omega=\sqrt{k/\mu}$. Returning to the original variables, we can write the work observable as $W_\mr{E\to S}(t_2,t_1)=K_1(t_2)-K_1(t_1)$, where $K_1=P_1^2/2m_1$ and
\be 
P_1(t)=a(t)\,P_1\sch+b(t)\,P_2\sch+c(t)\left(X_2\sch-X_1\sch\right),
\label{P1-elastic}
\ee 
\be
\begin{array}{l}
a(t)=\left[m_1+m_2\cos\left(\omega t\right)\right]/M,\\ 
b(t)=\left[1-\cos\left(\omega t\right)\right]m_1/M,\\ 
c(t)=\mu\omega\sin\left(\omega t\right).
\end{array}
\label{abc}
\ee
Although an explicit form of the work operator can be easily derived from the above formulas, it is not insightful and its diagonalization is not trivially feasible. We then restrict our analysis to convenient instants of time. An instructive example is
\be
W_\mr{E\to S}\left(v\tau,u\tau\right)=2\left[\tfrac{m_{1}-m_{2}}{M^{2}}P_1\sch P_2\sch+\tfrac{m_{1}}{M^{2}}(P_2\sch)^{2}-\tfrac{m_{2}}{M^{2}}(P_1\sch)^{2}\right],
\label{W21-elastic}
\ee
with $\tau=\pi/\omega$, $v$ ($u$) an odd (even) integer, and $v>u\geq 0$. Here, $P_1\sch P_2\sch$ is used as a shorthand for $P_1\sch\otimes P_2\sch$.  The work operator is diagonal in the composite basis $\{\ket{p_1}\ket{p_2}\}$, where $P_i\sch\ket{p_i}=p_i\ket{p_i}$, and its eigenvalues read 
\be 
w_{p_{1},p_{2}}\left(v\tau,u\tau\right)=2\left(\tfrac{m_{1}-m_{2}}{M^{2}}p_{1}p_{2}+\tfrac{m_{1}}{M^{2}}p_{2}^{2}-\tfrac{m_{2}}{M^{2}}p_{1}^{2}\right).
\label{wp1p2}
\ee 
Hence, by measuring $P_{1,2}\sch$ one prepares the amount $w_{p_1,p_2}$ of work in the time interval $[u\tau,v\tau]$. Notice, however, that this is a joint measurement, that is, both particles have to be touched. This example offers a better picture of the typical work operator: it is an observable acting on a composite vector space and quantum theory allows, in these cases, the existence of entangled work eigenstates. To see this, consider the initial state $\ket{\psi_0}=\ket{\bar{p}_1}\ket{\bar{p}_2}$, where $\ket{\bar{p}_{1,2}}$ are sharp Gaussian states centered at $\bar{p}_{1,2}$, thus ensuring a well defined mean work. Now, in terms of the relative and center-of-mass momentum, the work operator \eqref{W21-elastic} reads $W_\mr{E\to S}\left(v\tau,u\tau\right)=\frac{2}{M} P\cm\sch P\rr\sch$. This change of variables implies that
\be 
\frac{e^{-\frac{(p_1-\bar{p}_1)^2}{4\delta_p^2}}}{(2\pi\delta_p^2)^{1/4}}\frac{e^{-\frac{(p_2-\bar{p}_2)^2}{4\delta_p^2}}}{(2\pi\delta_p^2)^{1/4}} = \frac{e^{-\frac{(p\rr-\bar{p}\rr)^2}{4\delta\rr^2}}}{(2\pi\delta_p^2)^{1/4}}\frac{e^{-\frac{(p\cm-\bar{p}\cm)^2}{4\delta\cm^2}}}{(2\pi\delta_p^2)^{1/4}}e^{-\frac{\alpha p\rr p\cm+\beta}{\delta_p^2}},
\ee 
where $\bar{p}\rr$, $\bar{p}\cm$, $\delta\rr$, $\delta\cm$, $\alpha$, and $\beta$ are functions of the width $\delta_p$ and $\bar{p}_{1,2}$. From $\alpha=\frac{1}{2}\frac{m_2-m_1}{m_1+m_2}$ we see that the initial state is entangled in the new degrees of freedom whenever $m_2\neq m_1$. Conversely, preparing a product state $\ket{\bar{p}\rr}\ket{\bar{p}\cm}$ for these ``nonlocal degrees of freedom'' renders an  entangled state in $p_{1,2}$.

Recognizing that work is an observable allows one to envisage interesting scenarios whereby Alice can steer the work done in Bob's laboratory. Consider an instance in which an ancillary system ${\mr{A}}$ is prepared, along with particles 1 and 2, in the state $\ket{\Psi}=\alpha\ket{a}\ket{w_{p_1,p_2}}+\beta\ket{\bar{a}}\ket{w_{\bar{p}_1,\bar{p}_2}}$, where $\ket{w_{p_1,p_2}}\equiv \ket{p_1}\ket{p_2}$ and $\braket{a|\bar{a}}=0$. Let us assume that this state can be preserved while the ancilla is transported to Alice's laboratory, which is far away from Bob's site where particles 1 and 2 are allowed to interact with each other according to the Hamiltonian \eqref{H-elastic}. For generic amplitudes $\alpha$ and $\beta$, the state $\ket{\Psi}$ is entangled and possesses other subtle quantum correlations, such as Einstein-Podolsky-Rosen (EPR) steering \cite{Uola2020} and Bell nonlocality \cite{Brunner2014}. This does not imply that Alice can effectively control the average amount of work that will take place in Bob's laboratory, for her measurement outcomes are random. The presence of EPR steering means that one cannot ascertain, before any measurement, an element of reality for work. That is, referring back to Eq. \eqref{wp1p2}, the amount of work that particle 2 will do on particle 1 in the time interval $\left[u\tau,v\tau\right]$ cannot be ensured by any local hidden variable theory to be either $w_{p_1,p_2}$ or $w_{\bar{p}_1,\bar{p}_2}$. Only after Alice measures the ancilla in her far distant laboratory, thus collapsing the system state to either $\ket{a}\ket{w_{p_1,p_2}}$ or $\ket{\bar{a}}\ket{w_{\bar{p}_1,\bar{p}_2}}$, will the amount of work in Bob's laboratory materialize to either $w_{p_1,p_2}$ or $w_{\bar{p}_1,\bar{p}_2}$, respectively. What is more, such nonlocal work steering can be implemented by Alice's measurement at any instant after the state preparation at $t=0$ and prior to the time interval $\left[u\tau,v\tau\right]$.

It is worth noticing that one does not really {\it measure} a two-time observable, like work or displacement, at a given instant of time $t$ after the preparation. In fact, this cannot be done even within the classical paradigm simply because a two-time observable is not definable at a single time. Instead, we {\it prepare} it for the time interval $[t_1,t_2]$ through the establishment of $\rho_0$ at $t=0$. As we shall discuss later, work should accordingly be viewed as a two-time element of reality.

\subsection{Work statistics}
\label{sec:statistics}

Once work is treated as an observable, one can determine its eigenstates, the corresponding projectors, and the emergent probability distributions. In other words, one can raise the entire statistics associated with its preparation. An alternative view that has commonly been adopted, especially in contexts involving an external driving parameter $\lambda(t)$, conceives work as a stochastic variable that can only be inferred through a TPM protocol \cite{Campisi2011,Hanggi2017,Talkner2007}. The TPM scheme has enabled the experimental validation of important quantum fluctuation theorems (see, for instance, Refs. \cite{Campisi2011,Hanggi2017,Huber2008} and references therein) and it gives a relatively simple and fairly general way of accounting for work statistics in the quantum thermodynamics domain. In this section, we confront these two views emphasizing their conflicting results. 

We start with a brief review of the TPM protocol, which is often applied to a system $\mr{S}$ described by a time-dependent Hamiltonian $H_\mr{S}\sch(t)$. After being prepared at $t=0$ in a generic state $\rho_\mr{S}$, the system is submitted to a projective measurement of energy at $t_1$, thus jumping to an $H_\mr{S}\sch(t_1)$ eigenstate $\ket{e_n}$ with probability $\mf{p}_n=\braket{e_n|\rho_\mr{S}|e_n}$. The system then evolves unitarily (via $U_\dt$, with $\dt=t_2-t_1$) until the instant $t_2$, when a second measurement is performed and a random eigenvalue $\varepsilon_m$ of $H_{\mr{S}}\sch(t_2)$ is obtained with probability $\mf{p}_{m|n}=|\bra{\varepsilon_m} U_{\dt}\ket{e_n}|^2$. In this run of the experiment, work is computed as $w_{mn}=\varepsilon_m-e_n$. After many runs, the probability density associated with work is built as $\wp_w=\sum_{mn}\mf{p}_{m|n}\,\mf{p}_{n}\,\delta_{\tx{\tiny D}}(w-w_{mn})$, where $\delta_{\tx{\tiny D}}$ is the Dirac delta function and $\int dw\,\wp_w=1$. It follows that the $k$-th moment of work can be evaluated as $\overline{w^k} =\int dw\,\wp_w w^k= \sum_{mn}\mf{p}_{m|n}\mf{p}_{n}w_{mn}^{k}$. Notice that what is directly measured is energy, not work. In fact, a way to measure work without knowledge of the energy operator had not yet been presented until very recently \cite{Beyer2020}. Typically, the application of the TPM protocol to thermodynamic contexts presumes a nonautonomous dynamics wherein $H_\mr{S}\sch(t)=H_\mr{S}\sch\left(\lambda(t)\right)$, that is, the system ${\mr S}$ is externally controlled by a classical device whose dynamics is encoded in $\lambda(t)$ \cite{Campisi2011,Strasberg2021}. Moreover, the coupling with this device is regarded as part of the internal energy \cite{Huber2008}, which characterizes the inclusive approach \cite{Jarzynski2007}. Finally, it is usual to consider initial states such that $[\rho_\mr{S}(0),H_\mr{S}\sch(0)]=0$  \cite{Huber2008,Campisi2011,Hanggi2017}. Since these conditions are not met in the mechanical context under scrutiny here, some adaptations in the protocol will be in order.

Now, to appreciate the differences among the approaches, we focus on the problem of free fall, as treated in Sec. \ref{sec:gravity}. The corresponding classical model encompasses the Hamiltonian function $\mc{H}=p^2/2m+mgx$, which yields the work $\mc{W}_\mr{E\to S}=\Delta\mc{K}=-\Delta\mc{V}$, where $\mc{K}=p^2/2m$ and $\mc{V}=mgx$. This double identity, which also emerges in our approach to quantum work, induces one to conceive two variations of the typical TPM protocol, one for measurements of momentum (and thus kinetic energy) and another involving measurements of position (potential energy). Accordingly, we refer to these protocols as ${\mr{TPM}}_p$
and ${\mr{TPM}}_x$. In both cases, the preparation will be a Gaussian pure state $\rho_0=\ket{\psi_0}\bra{\psi_0}$ with corresponding wave function
\be
\psi_0(x)=\braket{x|\psi_0}=\left(2\pi\sigma_x^2\right)^{  -\frac{1}{4}}\exp\left[-\frac{(x-x_{0})^2}{4\sigma_x^2}+\frac{i p_{0}x}{\hbar}\right],
\label{psi0(x)}
\ee
where $\braket{X\sch}=x_0$, $\braket{P\sch}=p_0$, and $\sqrt{\braket{(X_*\sch)^2}}=\sigma_x$ with $X_*\sch\equiv X\sch-\braket{X\sch}$. For notation compactness, we use $\mc{G}_u\left(\bar{u},\sigma\right)=(2\pi\sigma^2)^{-\frac{1}{2}}\exp\left[-\frac{(u-\bar{u})^2}{2\sigma^2}\right]$ for Gaussian functions with width $\sigma$ and center at $\bar{u}$. The next step consists of unitarily evolving the initial state from $t=0$ until $t=t_1$ via the time-evolution operator \footnote{To obtain the factorized form \eqref{U_t1}, we first move the description to the interaction picture via the transformation $\ket{\overline{\psi}}=e^{i mgXt/\hbar}\ket{\psi}$. The time-evolution operator for the new Schr\"odinger equation,  $i \hbar \, \partial_t\ket{\overline{\psi}}=\frac{(P-mgt\mbb{1})^2}{2m}\ket{\overline{\psi}}$, can be directly factorized as $e^{-i P^2t/2m\hbar}\mr{e}^{i gt^2P/2\hbar}$.}
\be
U_{t_1}=e^{-i\Theta_{t_1}}\exp\left(-\frac{i mgt_1 X}{\hbar}\right)\exp\left(-\frac{i P^2t_1}{2m\hbar}\right)\exp\left(\frac{i gt_1^2 P}{2\hbar}\right),
\label{U_t1}
\ee 
where $\Theta_{t_1}=\frac{mg^2t_1^3}{6\hbar}$. The probability densities associated with the outcomes $r_i$ in measurements of $r\in\{x,p\}$ at $t_1$ read
\be\begin{array}{l}
\wp_{x_i}=\left|\braket{x_i|\psi(t_1)}\right|^2=\mc{G}_{x_i}\left(x_0+\tfrac{p_0t_1}{m}-\tfrac{gt_1^2}{2},\sigma_x(t_1)\right),\\
\wp_{p_i}=\left|\braket{p_i|\psi(t_1)}\right|^2=\mc{G}_{p_i}\left(p_0-mgt_1,\tfrac{\hbar}{2\sigma_x}\right),
\end{array}\label{wp_ri}
\ee
where  $\sigma_x(t_1)=\sigma_x\sqrt{1+(\hbar t_1/2m\sigma_x^2)^2}$. We assume that a position (momentum) measurement results in a Gaussian state with width $d_x$ $(d_p)$ referring to the measurement resolution (which are eventually taken to be ideal, that is, $d_{r}\to 0$). After the measurement is effectively performed, the state reduces to the eigenstate $\ket{r_i}$, here expressed as $\braket{r|r_i}=\sqrt{\mc{G}_r\left(r_i,d_r\right)}$.
Then, we apply the evolution operator \eqref{U_t1} with the change $t_1\to \dt=t_2-t_1$ and compute the conditional probability densities
\be\begin{array}{l}
\wp_{x_f|x_i}=\left|\braket{x_f|U_\dt|x_i}\right|^2=\mc{G}_{x_f}\left(x_i-\tfrac{g\dt^2}{2},d_x(\dt)\right),\\
\wp_{p_f|p_i}=\left|\braket{p_f|U_\dt|p_i}\right|^2=\mc{G}_{p_f}\left(p_i-mg\dt,d_p\right),
\end{array}
\label{wp_pf|pi}
\ee 
where $d_x(\dt)=d_x\sqrt{1+(\hbar \dt/2md_x^2)^2}$. Taking $e_{i,f}^{\text{\tiny $[x]$}}=mgx_{f,i}$ and $e_{i,f}^{\text{\tiny $[p]$}}=p_{i,f}^2/2m$ for the ${\mr{TPM}}_x$ and ${\mr{TPM}}_p$ protocols, respectively, we can compute work distributions as $\wp_w^{\text{\tiny $[r]$}}(t_2,t_1)=\int\!\!\int dr_i dr_f\,\wp_{r_i}\wp_{r_f|r_i} \delta_{\tx{\tiny D}}\left[w-(e_f^{\text{\tiny $[r]$}}-e_i^{\text{\tiny $[r]$}})\right]$. The results read
\be 
\wp_w^{\text{\tiny $[r]$}}(t_2,t_1)=\mc{G}_w\left(w^{\text{\tiny $[r]$}}(t_2,t_1),\sigma_w^{\text{\tiny $[r]$}}\right),
\label{wp_w}
\ee 
where $w^{\text{\tiny $[r]$}}(t_2,t_1)={\mr{f}}_g\Delta_x^{\text{\tiny $[r]$}}$ (mean work), ${\mr{f}}_g=-mg$, and
\be 
\begin{array}{lcl}
	\Delta_x^{\text{\tiny $[x]$}}=-\frac{g\dt^2}{2}, & \quad & \sigma_w^{\text{\tiny $[x]$}}=|{\mr{f}}_g|d_x(\dt), \\
	\Delta_x^{\text{\tiny $[p]$}}=\frac{p_0\dt}{m}-\frac{g(t_2^2-t_1^2)}{2}, & \quad & \sigma_w^{\text{\tiny $[p]$}}=\sigma_p g \dt.
\end{array}
\label{w_TPMr}
\ee 
The results for the ${\mr{TPM}}_p$ protocol were strictly derived  for $d_p\to 0$. The corresponding limit, $d_x\to 0$, for the ${\mr{TPM}}_x$ protocol would lead to $\sigma_w^{\text{\tiny $[x]$}}\to \infty$, which already points out a dramatic difference between these protocols. With the distributions \eqref{wp_w} we can compute the moment   $\overline{w^k}_{\text{\tiny $[r]$}}=\int dw\,\wp_w^{\text{\tiny $[r]$}}\,w^k$, and then the mean work $\overline{w}_{\text{\tiny $[r]$}}=\int dw\,\wp_w^{\text{\tiny $[r]$}}\,w=w^{\text{\tiny $[r]$}}(t_2,t_1)$ and the fluctuation $(\overline{w^2}_{\text{\tiny $[r]$}}-\overline{w}_{\text{\tiny $[r]$}}^2)^{1/2}=\sigma_w^{\text{\tiny $[r]$}}$. 

The differences are remarkable and symptomatic. In the ${\mr{TPM}}_x$ protocol, the mean momentum $p_0$ is erased by the first position measurement, thus leading to an average work $w^{\text{\tiny $[x]$}}$ whose associated displacement $\Delta_x^{\text{\tiny $[x]$}}=-\frac{g\dt^2}{2}$ can be very different from the one emerging in the ${\mr{TPM}}_p$ protocol, $\Delta_x^{\text{\tiny $[p]$}}=\frac{p_0\dt}{m}-\frac{g\dt^2}{2}$. Most importantly, in the limit as $d_x\to 0$ one has $\sigma_w^{\text{\tiny $[x]$}}/\sigma_w^{\text{\tiny $[p]$}}\approx\sigma_x/d_x\to \infty$, proving the huge differences implied by each protocol to the work fluctuation. Here we have an odd state of affairs because the two TPM protocols cannot agree about the mean work done on the particle. In other words, the TPM-based stochastic view fails to validate the identity $\overline{\mc{K}(t_2)-\mc{K}(t_1)}=\overline{\mc{V}(t_1)-\mc{V}(t_2)}$ in a conservative system.

As far as the work operator \eqref{Wg-DK} is concerned, one readily identify its eigenstates $\ket{p}$ and respective eigenvalues $w_p(t_2,t_1)={\mr{f}}_g\Delta x$, where $\Delta x=\frac{p_0\dt}{m}-\frac{g\dt^2}{2}=\Delta_x^{\text{\tiny $[p]$}}$. It then follows that the work distribution can be constructed via momentum distribution as
\be 
\int_{w_1}^{w_2}\!\!dw\,\wp_w=\int_{p_1}^{p_2}\!\!dp\,\left|\braket{p|\psi_0}\right|^2,
\ee 
with the change of dummy variables $w=-g\dt \,p+\frac{mg^2 \dt^2}{2}$ and $dw=-g\dt\,dp$. We then find
\be 
\wp_w=\mc{G}_w\left(w^{\text{\tiny $[p]$}},\sigma_w^{\text{\tiny $[p]$}}\right),
\ee 
from which we can compute the $k$-th moment of work, $\braket{w^k}=\int dw\,p_w w^k $, and, in particular, $\braket{w}={\mr{f}}_g\Delta x=w^{\text{\tiny $[p]$}}$ and $\sigma_w =(\braket{w^2}-\braket{w}^2)^{1/2}=\sigma_w^{\text{\tiny $[p]$}}$. We see from the above results that $\wp_w=\wp_w^{\text{\tiny $[p]$}}\neq \wp_w^{\text{\tiny $[x]$}}$, which shows that treating work either as an observable or as a stochastic variable whose value is revealed through TPM does not always give the same statistics. An important aspect contributing to the invalidation of the $\mr{TPM}_x$ protocol is that this approach gives a mean work inconsistent with the classical predictions, which are expected to be retrieved in light of the application of the Ehrenfest theorem to less-than-cubic potentials. Technically, the ``incompatibility'' of the $\mr{TPM}_x$ protocol with the other two approaches can be acknowledged by the fact that $[W_\mr{E\to S}(t_2,t_1),X(t)]=i \hbar g\dt$ whereas $[W_\mr{E\to S}(t_2,t_1),P(t)]=0$, relations that can also be rephrased with $V$ and $K$, respectively. In the latter case, measurements of the internal energy $E_\mr{S}=K$ (as demanded in the $\mr{TPM}_p$ protocol) are not able to disturb work states (which can be the initial states in our approach), so that the two approaches are expected to be compatible. However, in the former case, the position measurement introduces a huge momentum fluctuation, which ultimately leads to the elimination of the information about $p_0$ and $\sigma_x$ from the statistics. It is noteworthy that the inadequacy of TPM protocols---at least for the present context and purposes---is not exclusively related to the invasive nature of measurements. It also emerges from the assumption that joint probability distributions can be constructed in the form $\mf{p}_{m|n}\mf{p}_n$. This rationale is not valid in general quantum contexts, as we shall discuss later, although it has proven useful for thermodynamics approaches. 

To further illustrate the difficulties underlying the $\mr{TPM}_x$ approach, we consider the task of computing the mean instantaneous power associated with the gravitational field. Within the operator-based formalism, we can define the instantaneous power observable as $\mbb{P}(t)=\lim_{\tau\to 0}[E_\mr{S}(t+\tau)-E_\mr{S} (t)]/\tau$, which, in light of Eq. \eqref{Wg-DK}, results in
\be 
\mbb{P}(t)=\lim_{\tau}\frac{W_\mr{E\to S}(t+\tau,t)}{\tau}=-mg\left(\frac{P\sch}{m}-gt\right).
\ee 
We then readily obtains $\braket{\mbb{P}(t)}={\mr{f}}_g\left(\frac{p_0}{m}-gt\right)$, which is the traditional ``force $\times$ velocity'' statement of instantaneous power and the correct classical limit for the model under scrutiny. Now, resorting to the TPM-based formulas \eqref{w_TPMr} we find
\beq 
&&\overline{\mbb{P}}_{\mr{TPM}_x}(t)=\lim_{\tau \to 0}\frac{\overline{w}_{\tx{\tiny $[x]$}}(t+\tau,t)}{\tau}=0, \\
&&\overline{\mbb{P}}_{\mr{TPM}_p}(t)=\lim_{\tau \to 0}\frac{\overline{w}_{\tx{\tiny $[p]$}}(t+\tau,t)}{\tau}={\mr{f}}_g\left(\frac{p_0}{m}-gt\right).
\eeq 
Once again the differences are notorious: the $\mr{TPM}_x$ protocol is unable to reproduce the time dependence expected for the instantaneous power. An important symptom of the conflict already appears in the fact that the $\mr{TPM}_x$-based stochastic work is invariant under time translation. Mathematically, $\overline{w}_{\tx{\tiny [x]}}(t_2+\delta_t,t_1+\delta_t)=\overline{w}_{\tx{\tiny [x]}}(t_2,t_1)$. Work, however, is not supposed to be so, because it is not a state variable, that is, it should depend on the ``time path''. Indeed, we see that $W_\mr{E\to S}(t_2+\delta_t,t_1+\delta_t) \neq W_\mr{E\to S}(t_2,t_1)$. Even though the $\mr{TPM}_p$ protocol has incidentally produced results compatible with the operator-based formalism, this does not change the fact that the former approach conceives a sort of ``hybrid'' view of nature. That is, one assumes, on the one hand, that two-time physical quantities are stochastic variables not describable as quantum mechanical observables, but one does admit, on the other hand, that quantum mechanics can be applied to describe the deterministic time evolution between the contiguous measurements and the associated probability distributions. In contrast, our approach indicates how to accommodate two-time observables in the standard quantum structure.

Schemes for measuring work other than TPM protocols exist in which only one (generalized) quantum measurement is needed~\cite{Roncaglia2014,Beyer2020}. In general, though, energy measurements ``classicalize'' the system state, since they remove quantum coherence and make the energy become an element of reality \cite{Bilobran2015}. Most importantly, the very act of measuring is a relevant source of work and heat which, as such, must not be excluded from the energy balance \cite{Elouard2017,Perarnau2017}. The idea advocated here, that a Heisenberg model for two-time observables should be preferred in mechanical contexts in relation to $\mr{TPM}$ methods, is not restricted to the concept of mechanical work. In Appendix \ref{sec:Stoch-Obs}, we show that the same issues arise when one considers the free-particle displacement operator and the spin angular displacement as stochastic variables. Finally, however, it should ultimately be acknowledged that our results do not eliminate the operational success and the arguable adequacy of the TPM scheme in thermodynamics scenarios, where the proper conditions for the TPM method are fulfilled.

\subsection{Work as a two-time element of reality}
\label{sec:EoR}

In their 1935 paper~\cite{EPR1935}, Einstein, Podolsky, and Rosen (EPR) associated the notion of {\it element of reality} with the condition of full predictability of a given physical quantity. According to their criterion, ``if, without in any way disturbing a system, we can predict with certainty the value of a physical quantity, then there exists an element of physical reality corresponding to this physical quantity''. So, if a spin-$\frac{1}{2}$ particle is prepared in the state $\ket{0}=\left(\ket{+}+\ket{-}\right)/\sqrt{2}$, thus implying a null mean-square deviation for the $z$-component ($\sigma_{S_z}=0$), then $S_z$ is an element of reality, while $S_x$ is not ($\sigma_{S_x}>0$). Accordingly, $\ket{0}$ is said to be an $S_z$ state of reality\footnote{An alternative approach put forward by Bilobran and Angelo~\cite{Bilobran2015} claims that a state like $\mf{p}\ket{0}\bra{0}+(1-\mf{p})\ket{1}\bra{1}$ also implies an element of reality for $S_z$, even being such that $\sigma_{S_z}>0$. Although in this case there is a remaining unpredictability concerning this observable, it derives from a classical mixture of well-established elements of reality. For the sake of simplicity, throughout this work we restrict our analysis to the EPR elements of reality, leaving the discussion regarding this alternative view to be done elsewhere.}.

However tempting it might be, treating a two-time observable as a stochastic variable that becomes an EPR element of reality only via the realization of a TPM protocol actually dismisses the quantum subtleties underlying such an object and is, ultimately, unjustifiable from a fundamental viewpoint. To make this point thoroughly, we consider the two-time operator $C(t_2,t_1)=\frac{1}{2}\{A(t_1),B(t_2)\}$, with $A(t)=\sum_aa\Lambda_a(t)$ and $B(t)=\sum_bb\Lambda_b(t)$ denoting nondegenerate discrete-spectrum observables with respective Heisenberg projectors $\Lambda_{a,b}(t)=\phi_t(\Lambda_{a,b}\sch)$. Starting with the mean value $\braket{C}(t_2,t_1)=\Tr[C(t_2,t_1)\rho_0]$, one obtains
\be 
\braket{C}(t_2,t_1)=\sum_{a,b}ab\Tr[\Lambda_b(t_2)\Gamma_{a}(t_1)]\,\mf{p}(a,t_1),
\ee 
where 
\be
\Gamma_{a}(t_1)=\frac{\{\rho_0,\Lambda_a(t_1)\}}{2\mf{p}(a,t_1)}, \qquad \mf{p}(a,t_1)=\Tr[\Lambda_a(t_1)\rho_0].
\ee 
Because $\Gamma_a(t_1)\neq \Lambda_a(t_1)$, one can immediately conclude that $\Tr[\Lambda_b(t_2)\Gamma_{a}(t_1)]\neq \mf{p}(b,t_2|a,t_1)=\Tr[\Lambda_b(t_2)\Lambda_a(t_1)]$ and $\braket{C}(t_2,t_1)\neq \sum_{a,b}ab\,\mf{p}(b,t_2|a,t_1)\mf{p}(a,t_1)$. Moreover, even though $\Gamma_a^\dagger=\Gamma_a$ and $\Tr(\Gamma_a)=1$, one can prove that $\Gamma_a$ is not semi-positive definite in general and, therefore, is not a quantum state. A simple illustration follows with $\rho_0=\ket{0}\bra{0}$, $\Lambda_a(t_1)=\ket{+}\bra{+}$, and $\ket{\theta}=\cos{\theta}\,\ket{0}+\sin{\theta}\,\ket{1}$, for a two-level system. We then find $\braket{\theta|\Gamma_a(t_1)|\theta}=\cos\theta\,(\cos\theta+\sin\theta)$, which is negative for $\theta\in\left(\frac{\pi}{2},\frac{3\pi}{4}\right)$. This shows that conceiving the statistics of the two-time operator $C(t_2,t_1)$ as emerging from the direct product of individual elements of reality $a$ and $b$ weighted by some tentative joint probabilities, such as $\Tr[\Lambda_b(t_2)\Gamma_{a}(t_1)]\mf{p}(a,t_1)$ (a pseudo joint probability) or $\mf{p}(b,t_2|a,t_1)\mf{p}(a,t_1)$ (the TPM prescription), may be a hasty move in general. Indeed, as we have already seen in Sec.~\ref{sec:statistics}, a construction like $\mf{p}(b,t_2|a,t_1)\mf{p}(a,t_1)$ may fail to validate energy conservation. Furthermore, we should recall that the covariance function $\frac{1}{2}\braket{\{A(t_1),B(t_2)\}}-\braket{A(t_1)}\braket{B(t_2)}$ and a related witnesses of ``time nonlocality'' \cite{Leggett1985} already warn us about the idea of statistics factorability in time.

Thanks to classical mechanics and statistical physics, we got used to the idea that the whole universe is described by a local-in-time state of affairs, with the position and momentum of each particle being elements of reality at each instant of time (realism). Accordingly, one might reject the displacement $x(t_2)-x(t_1)$ of a particle as an element of reality because its constituents $x(t_{1,2})$ do not belong to the same time-local reality state. This view, however, is too restrictive. There is no problem in viewing $x(t_2)-x(t_1)$ as a two-time element of reality, since it is a relevant physical concept that becomes fully defined as long as one specifies two instants of time. Accordingly, the notion of mean velocity emerges as a relevant element of reality as well, and so does work. This is not to say, however, that the entire segment $x(t_2)-x(t_1)$ of trajectory is an element of reality, although this would be admissible in classical physics. The statement is weaker, as it focuses only on those two specific instants of time. Moreover, there is no need for one to consider displacement as a combination of two elements of reality; once we fix $t_{1,2}$, displacement can be viewed as an ``indivisible quantity'' (a unit {\it per se}), with an element of reality $p_0(t_2-t_1)/m$. 

Referring back to the operator $C(t_2,t_1)=\frac{1}{2}\{A(t_1),B(t_2)\}$, one may regard $C(t_2,t_1)$ as an observable in its own right, with its own elements of reality, and not as a mere mix of the observables $A(t_1)$ and $B(t_2)$. To this end, we have to fix $t_{1,2}$, express $C(t_2,t_1)$ in terms of Schr\"odinger operators, and then diagonalize it for these specific instants of time. Having obtained $C(t_2,t_1)\Lambda_c^i(t_2,t_1)=c\Lambda_c^i(t_2,t_1)$ (with eventual degeneracy $i=1,2,\cdots,g_c$), we can think of two-time elements of reality $c$, not necessarily equal to $ab$, with respective two-time probabilities $\mf{p}(c,t_2,t_1)=\sum_i\Tr[\Lambda_c^i(t_2,t_1)\,\rho_0]$. The preparation $\Lambda_c^i(t_2,t_1)$ guarantees a well defined value for $C(t_2,t_1)$ for the fixed times $t_{1,2}$, that is, a two-time element of reality. This preparation, however, does not ensure that there will be an element of reality for $C(t_4,t_3)$ with $t_{4,3}\neq t_{1,2}$ because $\Lambda_c^i(t_2,t_1)$ will not necessarily be an eigenstate of $C(t_4,t_3)$. As we have illustrated in previous sections, the work operator perfectly fits in this picture, since we can always obtain the eigenvalue relation $W_\mr{E\to S}(t_2,t_1)\Lambda_w^i (t_2,t_1)=w(t_2,t_1)\Lambda_w^i (t_2,t_1)$ for some work projectors $\Lambda_{w}^i(t_2,t_1)$. In Secs.~\ref{sec:gravity} and \ref{sec:quadratic}, we have found $\Lambda_w(t_2,t_1)=\ket{p}\bra{p}$ and $\Lambda_w^i(v\tau,u\tau)=\ket{p_1}\bra{p_1}\otimes\ket{p_2}\bra{p_2}$ (with high degeneracy), respectively. We remind the reader that these constructions regarded momentum as a discrete variable. In a continuous-variable description, we can think of infinitesimal projectors like $d\Lambda_w(t_2,t_1)=\ket{p}\bra{p}dp$. This is the approach we employ in what follows.

\subsection{A Schr\"odinger-like picture for work}
\label{sec:SchPict}

Arguably, work stands out among the two-time observables because of its fundamental role in the law of conservation of energy. Here we point out another special facet of this concept within the quantum formalism. To this end, it is convenient to rephrase Eq.~\eqref{QM-Wji} in terms of the Heisenberg power $\mbb{P}(t)=\phi_t(\mbb{P}\sch)$, where $\mbb{P}\sch=\phi_t^*(-\frac{1}{2}\{\dot{X}_i,\partial_{X_i}V_{ij}\})$ is a function of Schr\"odinger operators. Then, we can write $W_{j\to i}(t_2,t_1)=\mbb{P}(t_2,t_1) \dt$ with the average power
\be 
\mbb{P}(t_2,t_1):=\int_{t_1}^{t_2}\!\frac{dt}{\dt}\,\phi_t(\mbb{P}\sch)\equiv \Phi_{t_2,t_1}(\mbb{P}\sch),
\label{Power}
\ee 
for $\dt=t_2-t_1$. Here we have introduced the time-averaging map $\Phi_{t_2,t_1}$ satisfying $\lim_{t_2\to t_1}\Phi_{t_2,t_1}=\phi_{t_1}$ and $\Phi_{t_2,t_1}=\Phi_{-t_2,-t_1}^*$. Of course, not all two-time observables admit a description in terms of this map. Let us consider the continuous-variable spectral decomposition $\mbb{P}\sch=\int\!d\Lambda_\pi\,\pi$, with orthogonal projectors $d\Lambda_\pi=\Lambda_\pi d\pi$ such that $\Lambda_\pi=\ket{\pi}\bra{\pi}$,  $d\Lambda_{\pi}d\Lambda_{\pi'}=\delta_{\pi,\pi'}d\Lambda_{\pi}$, $\int\!d\Lambda_\pi=\mbb{1}$, and $\Tr(d\Lambda_\pi)=1$. In the jargon introduced previously, $\phi_t(d\Lambda_\pi)$ is a state of reality (an eigenstate) associated with the power $\mbb{P}(t)=\int \phi_t(d\Lambda_\pi)\pi$. Note that a similar construction applies to all Heisenberg operators. Given the preparation $\rho_0$, the probability of finding the element of reality $\pi$ for the power $\mbb{P}\sch$ is $d\mf{p}(\pi)=\Tr(d\Lambda_\pi\,\rho_0)$. The work operator now reads
\be 
W_{j\to i}(t_2,t_1)=\int \Phi_{t_2,t_1}\left(d\Lambda_\pi\right)\,\pi\dt
\label{W-EoR}
\ee
and the probability associated with the element of reality $\pi\dt$ is given by
\be 
d\mf{p}(\pi,t_2,t_1)=\Tr\left[\rho_0\,\Phi_{t_2,t_1}\left(d\Lambda_\pi\right)\right].
\label{dp}
\ee 
This is a genuine probability since it is nonnegative and $\int d\mf{p}(\pi,t_2,t_1)=1$. Now we come to the crux: in light of the analogy $\Phi_{t_2,t_1}\rightleftharpoons \phi_t$, the structure of the relations \eqref{W-EoR} and \eqref{dp} becomes closely related to Heisenberg's formalism for one-time operators, like power. This view is strengthened by the fact that we can readily check the validity of the completeness relation $\int \Phi_{t_2,t_1}\left(d\Lambda_\pi\right)=\mbb{1}$ and the Schr\"odinger-like formulation $d\mf{p}(\pi,t_2,t_1)=\Tr\left[\Phi_{t_2,t_1}^*(\rho_0)\,d\Lambda_\pi\right]$. Likewise, the mean work
\be
\braket{W_{j\to i}(t_2,t_1)}\!=\!\!\int\!\! d \mf{p}(\pi,t_2,t_1)\,\pi\dt=\Tr\left[\rho_0\Phi_{t_2,t_1}\left(\mbb{P}\sch\dt\right)\right]
\label{MeanWork}
\ee
also admits a Schr\"odinger-like formulation in terms of the state 
\be 
\Phi_{t_2,t_1}^*(\rho_0)=\int_{t_1}^{t_2}\frac{dt}{\dt}\phi_t^*(\rho_0).
\label{mean_rho}
\ee 
Note that this is a legitimate quantum state, meaning that it is normalized, Hermitian, and semi-positive definite. Furthermore, it shows us the laborious way\footnote{In section \ref{sec:statistics}, we have shown how to build the work probability distribution starting with the expansion of the Heisenberg work operator in terms of a resulting Schr\"odinger operator, that is, $W_\mr{E\to S}(t_2,t_1)\equiv O\sch$. With that, the task is directly accomplished through the construction of the probability distribution associated with $O\sch$. In general, however, this procedure can be very laborious as well, for $O\sch$ is expected to be a joint observable demanding nonlocal measurements.} through which one can raise the distribution \eqref{dp}: in principle, one can determine $\rho(t)=\phi_t^*(\rho_0)$ via quantum state tomography at every instant of time and then compute the average theoretically. An important consequence of such ``time mixture'' is that the purity of $\Phi_{t_2,t_1}^*(\rho_0)$ is in general smaller than $\rho_0$'s. To explicitly verify this, we use the concavity and the unitary invariance of the von Neumann entropy $S$, which are respectively written as $S\left(\sum_ip_i\rho_i \right)\geqslant \sum_ip_iS(\rho_i)$ and $S\left(U\rho_i\, U^\dag\right)=S(\rho_i)$ for generic states $\rho_i$, probability distributions $p_i$, and unitary transformations $U$. With that, we arrive at $S(\Phi_{t_2,t_1}^*(\rho_0))\geqslant S(\rho_0)$, with equality holding if, and only if, $\Phi_{t_2,t_1}^*(\rho_0)=\rho_0$ (this occurs, for instance, when $\rho_0$ is a stationary state). It follows that $\Phi_{t_2,t_1}^*(\rho_0)$ typically is a mixed state. This is, however, no different from what we have obtained previously. Referring back to Eq.~\eqref{phi(dE)}, for instance, we see that the mean energy $\Tr\left[\rho_0\phi_{t_i}(H_\mr{S}\sch\otimes \mbb{1}_{\mr{E}})\right]$ equals $\Tr_\mr{S}\left[\Omega(t_i) H_\mr{S}\sch\right]$, where $\Omega_i=\Tr_{\mr{E}}[\phi_{t_i}^*(\rho_0)]$ is, in light of the Stinespring theorem, a mixed state. 

The take-away message is as follows. Generic two-time observables cannot be defined in terms of a genuine Schr\"odinger picture because there exists no sensible notion of two-time state. (Note that the scenario is equivalent in classical statistical physics.) Yet, quantum mechanical work does admit a formulation in terms of a Schr\"odinger-like picture, which, however, demands the notion of a time-mixture state. Its definition, Eq. \eqref{mean_rho}, clearly implements the time averaging originally codified in the definition of work in Heisenberg's picture. Besides attesting to the overall consistency of our approach, it suggests another way of performing tests of principle in the laboratory.

There is at least one point where the analogy with Heisenberg's formalism for one-time operators does not always apply: Eq.~\eqref{W-EoR} is not generally equivalent to the spectral decomposition of work in their states of reality. This is so because $\Phi_{t_2,t_1}(d\Lambda_\pi)$ is not necessarily a $W_{j\to i}(t_2,t_1)$ eigenstate. Nevertheless, this fact does not invalidate the Schr\"odinger-like view developed above, because Eqs.~\eqref{W-EoR}-\eqref{mean_rho} were derived without this assumption. On the other hand, when this condition is fulfilled, a preparation like $\rho_0=\Phi_{t_2,t_1}(d\Lambda_{\bar{\pi}})$ satisfies $W_{j\to i}(t_2,t_1)\rho_0=\bar{\pi}\dt\rho_0$. Hence, $\braket{W_{j\to i}(t_2,t_1)}=\bar{\pi}\dt$ and $\braket{W_{j\to i}^k(t_2,t_1)}=\braket{W_{j\to i}(t_2,t_1)}^k$, meaning that no quantum indefiniteness whatsoever can be associated with work. In this case, work becomes a two-time element of reality.

\subsection{Work-energy uncertainty principle}
\label{sec:uncertainty}

In connection with the previous discussion, here we emphasize  that once work is acknowledged as an observable, then one has to abandon the view that it is a mere combination of elements of reality. More importantly, quantum mechanics allows work to be an element of reality even when its constituents terms are not. This can be appreciated by use of uncertainty relations. Let $\sigma_O=\tx{\small $\sqrt{\braket{O_*^2}}$}$, with $O_*\equiv O-\braket{O}$, be the quantum uncertainty of a given Heisenberg operator $O$, and $W_{21}=H_2-H_1$ is a shorthand for the resultant work operator $W_\mr{E\to S}(t_2,t_1)$ given in Eq.~\eqref{phi(dE)}, where $H_i\equiv H_\mr{S}(t_i)$. Via the weak version of the uncertainty principle, we have $\sigma_{W_{12}}\sigma_{H_i}\geq \frac{1}{2}|\braket{[W_{21},H_i]}|$. Since $[W_{21},H_1]=[H_2,H_1]=[W_{21},H_2]$, one finds
\be
\sigma_{W_{21}}\left(\sigma_{H_1}+\sigma_{H_2}\right)\geq \left|\braket{[H_1,H_2]}\right|.
\label{UR}
\ee 
This (two-time) uncertainty relation shows that, excluding very particular states like $\rho_0=\mbb{1}/d$, with $d=\dim(\mbb{H}_\mr{S}\otimes \mbb{H}_{\mr{E}})$, which implies $\braket{[H_1,H_2]}=0$, the general situation is such that whenever the energy operators do not commute at different times, one cannot produce states ensuring arbitrarily small uncertainties for work and the initial and final energies simultaneously\footnote{Of course, relation \eqref{UR} is not an exclusiveness of the pair work-energy. It is easy to show that for any operator $\Gamma_{21}=A_1+A_2$, one can prove that $\sigma_{\Gamma_{21}}(\sigma_{A_2}+\sigma_{A_1})\geq |\braket{[A_1,A_2]}|$. In Sec.~\ref{sec:gravity}, an example is given for the problem regarding the displacement of a free particle. By preparing the system in a momentum eigenstate, we have $\sigma_{\delta_{21}}\to 0$ whereas $\sigma_{X_i}\to \infty$ (see also the examples provided in Appendix \ref{sec:Stoch-Obs}).}. A direct illustration  of this result was presented in Sec.~\ref{sec:gravity}, where via the preparation of a momentum eigenstate we had $\sigma_{W_{21}}\to 0$ and $\sigma_{V_i}\to \infty$, with $V_i$ being the gravitational potential energy and $[V_1,V_2]=i \hbar \dt/m$. In this case, work due to the gravitational potential is an element of reality, while the energies at times $t_{1,2}$ are not. Rather than seeing all this as a drawback for the notion of work operator \cite{Allahverdyan2014}---an attitude that is ultimately based on classical lines of thought--- here we defend that this actually is a subtle manifestation of a truly quantum notion of work.

Interestingly, for the aforementioned gravitational model, given $t_{1,2}$ one can always have a preparation $\ket{p_{12}}$, with specific value $p_{12}=mg(t_1+t_2)/2$, such that $w_{p_{12}}(t_2,t_1)=0$. In this case, energy conservation turns out to be an element of physical reality, that is, 
\be 
\braket{W_{21}}=\sigma_{W_{21}}=0 \qquad \text{(energy conservation).}
\label{W=sigma=0}
\ee 
As far as the quadratic potential is concerned, many regimes can be found where energy conservation emerges as part of the physical reality. Let us set, for instance, $t_1=\pi/(2\omega)$ and $t_2=3\pi/(2\omega)$ in Eqs. \eqref{abc} and \eqref{P1-elastic}. It then follows that $W_{21}=-\frac{2\mu\omega}{M} P\cm X\rr$, from which we see that a preparation like $\ket{p\cm=0}\ket{x\rr}$ ($\forall x\rr$)---corresponding to an entangled state in the laboratory coordinates---guarantees $\braket{W_{21}}=\sigma_{W_{21}}=0$. In this case, even though the $\mr{S}$ energy is an element of reality at the instants $t_{1,2}$, the $\mr{S+E}$ energy is not, since the preparation is not an eigenstate of the total Hamiltonian. That is, the total energy is conserved but it is not an element of reality at the instants of time $t_{1,2}$. Yet, the conservation of the total energy is an element of reality.

This result is not specific to the model studied. Consider a generic autonomous scenario in which $\mr{S}$ interacts with ${\mr{E}}$ through some potential $V$. Let $H=H_\mr{S}+H_\mr{ext}$ be the pertinent time-independent Hamiltonian, with $H_\mr{ext}=H_{\mr{E}}+V$ being the external energy. The fact that $\Delta H=0$ directly implies, for any preparation $\rho_0$, that the conservation of the total energy is an element of the physical reality, that is
\be 
\braket{\Delta H}=\sigma_{\Delta H}=0 \qquad \text{(total energy conservation)}.
\ee 
However, since $\rho_0$ needs not be an eigenstate of $H$, it follows that, although surely conserved in time, the total energy is not readily implied to be an element of reality. Now, suppose in addition that $\rho_0$ is an eigenstate of $\Delta H_\mr{S}$ (and hence of the work done by ${\mr{E}}$ on $\mr{S}$). Because $\Delta H_\mr{ext}=-\Delta H_\mr{S}$, one has $\sigma_{\Delta H_\mr{ext}}=\sigma_{\Delta H_\mr{S}}=0$. Therefore, for conservative systems, if work is an element of reality, then the energy change of the rest of the universe will also be. This further illustrates the formal ground on which the law of conservation of energy is established within our approach.

\section{Concluding remarks}
\label{sec:conclusions}

The claim ``work is not an observable'' \cite{Talkner2007} has been derived from the premise that fluctuation theorems are correct. This is of course a good working hypothesis in several contexts, but it is debatable whether one should stick to this perspective in order to have a fundamental theory of microscopic mechanical systems. Should we add to the axiomatic structure of quantum mechanics the assumption that two-time variables, such as work, displacement, and velocity, are to be treated as stochastic variables? Here we answered this question in the negative for a genuinely quantum mechanical context involving  conservative autonomous few-particle systems. Not only did we show that work can be technically viewed as a quantum observable, but gave some examples in which the TPM-based stochastic view greatly deviates from the expected semiclassical results. Our results do not eliminate the operational relevance of the stochastic view for general contexts, but they show that it is not unavoidable from a fundamental viewpoint.

Our proposal for the concept of quantum mechanical work requires an updating of the common view that physical concepts refer to a time-local state of affairs. In fact, this cannot be the case even within the classical paradigm. As we have shown, once we conceive that a two-time observable can be described as a Heisenberg operator, with its fixed-time expansion in terms of Schr\"odinger operators and its related eigenbasis, then we learn how to derive the corresponding statistics, without in any way resorting to non-quantum methods. The whole characterization of the concept as a genuine quantum entity thus follows: the work done on a system (i) has a well-defined spectrum (eventually quantized), (ii) can be prepared as an element of reality via specific measurements, (iii) has an intrinsic quantum fluctuation, (iv) satisfies an uncertainty relation with energy, and (v) allows one to identify scenarios in which the conservation of energy can be guaranteed as an element of reality.

We hope that our findings may help to relax the skepticism concerning the adequacy of purely quantum mechanical descriptions of work (and other two-time physical quantities) and encourage researchers to explore the subtleties deriving from this concept. Another interesting route consists of extending the notion of work observable to other domains, such as the thermodynamics one. Studies along these lines are now in progress in our research group. 

\section*{Acknowledgments}
It is a pleasure to acknowledge fruitful conversations with Frederico Brito, Gabriel Landi, Marcus Bonan\c{c}a, Roberto Serra, Alexandre Orthey, Danilo Fucci, and Marcelo Janovitch. We also thank Sebastian Deffner and Eric Lutz for drawing our attention to the inclusive-versus-exclusive dilemma and the fundamental differences underlying the concept of work in mechanics and thermodynamics, respectively. Financial support from the National Institute for Science and Technology of Quantum Information (CNPq, INCT-IQ 465469/2014-0) and the Brazilian funding agency CNPq, under Grants No. 146434/2018-8 (T.A.B.P.S.) and No. 309373/2020-4 (R.M.A.), is gratefully acknowledged.
%
%
\appendix
\section{Quantum observables versus stochastic variables}
\label{sec:Stoch-Obs}

Here we provide further illustrations on the differences in treating time variations of physical quantities either as observables or as stochastic variables.

\subsection{Free-particle displacement}
\label{sec:displacement}

In this analysis, we follow the formalism used in Sec.~\ref{sec:statistics}. Let us define the displacement operator of a free particle as $\delta(t_2,t_1)=X(t_2)-X(t_1)=P\sch\dt/m$, where $m$ is the mass of the particle and $\dt=t_2-t_1$. The preparation is assumed to be a Gaussian pure state $\rho_0=\ket{\psi_0}\bra{\psi_0}$ with
\be
\psi_0(x)=\braket{x|\psi_0}=\left(2\pi\sigma_x^2\right)^{  -\frac{1}{4}}\exp\left[-\frac{(x-x_{0})^2}{4\sigma_x^2}+\frac{i p_{0}x}{\hbar}\right],
\label{psi0(x)}
\ee
where $\braket{X\sch}=x_0$, $\braket{P\sch}=p_0$, $\sqrt{\braket{(X_*\sch)^2}}=\sigma_x$, and $X_*\sch\equiv X\sch-\braket{X\sch}$. Again, we assume that a position measurement results in a Gaussian state with width $d_x$ (referring to the measurement resolution), and we use $\mc{G}_u\left(\bar{u},\sigma\right)=(2\pi\sigma^2)^{-\frac{1}{2}}\exp\left[-\frac{(u-\bar{u})^2}{2\sigma^2}\right]$ for Gaussian functions with width $\sigma$ and center at $\bar{u}$.

Treating displacement as a stochastic variable implies using a TPM protocol for its determination. The probability density of finding the value $x_i$ in a measurement of $x$ at $t_1=0$ reads
\be
\wp_{x_i}=\left|\braket{x_i|\psi_0}\right|^2=\mc{G}_{x_i}\left(x_0,\sigma_x\right).
\label{wp_xi}
\ee
After the measurement is effectively performed, the state reduces to the eigenstate $\ket{x_i}$, expressed as $\braket{x|x_i}=\sqrt{\mc{G}_{x}\left(x_i,d_x\right)}$. The next steps consist of unitarily evolving the resulting state with the unitary operator $U_t=e^{-i Ht/\hbar}$, where $H=P^2/2m$, and then computing the conditional probability density
\be\begin{array}{l}
	\wp_{x_f|x_i}=\left|\braket{x_f|U_t|x_i}\right|^2=\mc{G}_{x_f}\left(x_i,d_x(t)\right),
\end{array}
\label{wp_xf|xi}
\ee 
where $d_x(t)=d_x\sqrt{1+(\hbar t/2md_x^2)^2}$. The displacement distribution is given by $\wp_\delta^{\tx{\tiny TPM}}=\int\!\!\int d x_i d x_f\,\wp_{x_i}\wp_{x_f|x_i} \delta_{\tx{\tiny D}}\left[\delta-(x_f-x_i)\right]$, where $\delta_{\tx{\tiny D}}$ is the Dirac delta function.  The result reads
\be 
\wp_\delta^{\tx{\tiny TPM}}=\mc{G}_\delta\left(0,d_x(t)\right),
\label{wp_delta}
\ee 
implying null mean displacement and fluctuation $d_x(t)$. Now, as far as displacement is treated as the observable $\delta(t,0)=P\sch t/m$, it is obvious that its probability distribution is directly related to the momentum's, that is,
\be 
\wp_\delta=\mc{G}_\delta\left(\sfrac{p_0 t}{m},\sfrac{\sigma_ pt}{m}\right),
\ee 
which yields the mean displacement $p_0 t/m$ and the uncertainty $\sigma_\delta\equiv \sigma_p t/m=\hbar t/2m\sigma_x$. The differences between the two approaches are remarkable. In particular, for the ideal resolution $d_x\to 0$ one has $d_x(t)\cong \hbar t/2m d_x\to \infty$, showing that we completely loose the sense of displacement with the TPM statistics. In addition, it is clear that the first measurement in the TPM protocol dissipates any dependence of the result on the initial momentum $p_0$ and the width $\sigma_x$. In contrast, treating displacement as  a quantum observable yields an uncertainty $\hbar t/2m\sigma_x$, which can be significantly small for large values of $\sigma_x$ and/or $m$. However counter-intuitive it may sound---for one does not expect to find a well defined value of displacement when the positions at $0$ and $t$ are completely random---this actually is a fingerprint of two-time quantum observables. In this regard, trajectory-based interpretations of quantum mechanics may help us to make this point. Considering, for instance, the predictions of Bohmian mechanics, the trajectories for the system under scrutiny read \cite{Pan2010}
\be 
x(t)=x_0+\sfrac{p_0t}{m}+\left[x(0)-x_0\right]\text{\small $\sqrt{1+\left(\sfrac{\hbar t}{2m\sigma_x^2}\right)^2}$},
\ee 
where $x(0)$ defines the initial condition of each particular Bohmian trajectory. In this approach, these trajectories flow side-by-side governed by the quantum potential. Introducing the Ehrenfest time scale $t_E=2m\sigma_x^2/\hbar$ and considering the long-time regime, we find the displacement
\be 
x(t)-x(0)\cong \left(\sfrac{p_0}{m}+\sfrac{[x(0)-x_0]}{t_E}\right)t,
\ee 
which approaches $p_0t/m$ when $t_E$ is large enough. Therefore, within the Bohmian perspective, it is perfectly fine for a system to have a well defined displacement $(\sigma_\delta\to 0)$ and a highly uncertain position $(\sigma_x\to \infty)$. Interestingly, here we explicitly have the uncertainty relation $\sigma_\delta\sigma_x=\hbar t/2m$.

\subsection{Spin precession}
\label{sec:precession}

We now provide an illustration involving discrete variables. Consider the task of assessing the variation of the $y$-component of a spin-$\frac{1}{2}$ angular momentum in the dynamics imposed by $H=\omega S_z$ over the time interval $\big[0,\frac{\pi}{2\omega}\big]$, where $\omega$ is the precession frequency. Given the Heisenberg operator $S_y(t)=S_x\sch\sin{\omega t}+S_y\sch\cos{\omega t}$, it follows that treating the mentioned quantity as observable results in 
\be 
\delta S_y=S_y\left(\pi/2\omega\right)-S_y\left(0\right)=S_x\sch-S_y\sch\stackrel{.}{=}\tfrac{\hbar}{2}\left(
\begin{array}{cc} 0 & 1+i \\ 1-i & 0 \end{array}
\right),
\ee 
with eigenvalues $\epsilon\hbar/\sqrt{2}$ and respective eigenvectors $\ket{u_\epsilon}=\left(\ket{0}+\epsilon\,e^{-i \pi/4}\ket{1}\right)/\sqrt{2}$, where $\epsilon=\pm 1$. Let us consider the initial state $\ket{\psi_0}=\ket{+}=(\ket{0}+\ket{1})/\sqrt{2}$. The probability distribution for $\delta S_y$ then reads
\be 
\mf{p}_{\delta S_y}(\epsilon)=|\braket{u_\epsilon|\psi_0}|^2=\tfrac{1}{2}\left( 1+\tfrac{\epsilon}{\sqrt{2}}\right).
\label{p^obs}
\ee 
It is noteworthy that to obtain this distribution experimentally one needs to measure a single observable, $S_x\sch-S_y\sch$, right after the preparation of the pure ensemble $\ket{\psi_0}$. 
Let us now consider the stochastic-variable approach. The probability of getting $\epsilon\hbar/2$ in a measurement of $S_y$ at $t=0$ is given by $\mf{p}(\epsilon)=|\braket{y_\epsilon|\psi_0}|^2=1/2$, where $\ket{y_\epsilon}=\left(\ket{0}+i \epsilon\ket{1}\right)/\sqrt{2}$. The post-measurement state evolved until an instant $t>0$ reads $\ket{\psi(t)}=\left(e^{-i \omega t/2}\ket{0}+i \epsilon\,e^{i \omega t/2}\ket{1}\right)/\sqrt{2}$, which for the specific time $t=\pi/2\omega$ reduces to $\ket{\psi(\frac{\pi}{2\omega})}=\left(\ket{0}- \epsilon\ket{1}\right)/\sqrt{2}$. The probability of getting $\epsilon'\hbar/2$ in a subsequent measurement of spin is $\mf{p}(\epsilon'|\epsilon)=|\braket{y_{\epsilon'}|\psi(\frac{\pi}{2\omega})}|^2=1/2$. The resulting probability distribution is
\be 
\mf{p}_{\delta S_y}^{\tx{\tiny TPM}}=\mf{p}(\epsilon'|\epsilon)\mf{p}(\epsilon)=1/4.
\label{p^TPM}
\ee 
The divergence between the distributions \eqref{p^obs} and \eqref{p^TPM} is clear. In particular, with them we find the following mean values:
\beq 
&&\braket{\delta S_y}= \sum_{\epsilon}\left(\tfrac{\epsilon\hbar}{\sqrt{2}}\right)\mf{p}_{\delta S_y}(\epsilon)=\hbar/2,\\ 
&&\braket{\delta S_y}^{\tx{\tiny TPM}} =\sum_{\epsilon,\epsilon'}\left(\tfrac{\epsilon'\hbar}{2}\tfrac{\epsilon\hbar}{2}\right)\mf{p}_{\delta S_y}^{\tx{\tiny TPM}}=0.
\eeq 
We see that the TPM approach predicts null variation for the $y$-component of spin in the considered time interval, which is unexpected from a semiclassical perspective.


\end{document}